\documentstyle[prl,aps]{revtex}
\begin{document}
\draft

\title{\bf ENTANGLED STATES AND LOCAL MEASUREMENTS}

\author{Alexander A. Klyachko and Alexander S. Shumovsky}

\address{Faculty of Science, Bilkent University, Bilkent, Ankara,
06533, Turkey}

\maketitle

\begin{abstract}
We show that the maximum entanglement in a composite system
corresponds to the maximum uncertainty and maximum correlation of
local measurements.
\end{abstract}

\pacs{PACS number(s): 03.65.Ud, 03.67.-a}

\narrowtext

\twocolumn

\newpage

It has been recognized that entanglement is one of the most
symbolic, suggestive, and, at the same time, enigmatic concepts in
quantum mechanics. On the one hand this concept touches on the
fundamental notions of locality and reality in quantum physics
\cite{1,2,3} and on the other hand plays a key role in quantum
information theory \cite{4}.

The physical essence of entanglement consists in the existence of
quantum correlations between the individual parts of a composite
system that have interacted once in the past but are no longer
interacting. Formally, these correlations are caused by the
combination of the superposition principle in quantum mechanics
with the tensor product structure of the space of states \cite{5}.

In the usual treatment, the entanglement is associated with the
nonseparability of states of a composite quantum system. At the
same time, the nonseparability of states  by itself is not a
sufficient condition of either maximum or distilled entanglement
\cite{6}. We also note that the violation of Bell's inequalities
and other conditions of classical realism, which has attracted a
great deal of interest in the context of entanglement (e.g., see
\cite{3,4,7} and references therein), demonstrate nothing but
nonexistence of hidden variables in quantum mechanics and
therefore cannot be considered as a criterion of entanglement.
Moreover, there are a certain examples of mixed states \cite{8},
which do not violate Bell's inequalities but manifest entanglement
and therefore can be used for quantum teleportation \cite{9}.

Definitely the nonseparability of entangled states has a certain
informational aspect \cite{10}. In particular, there is an
assumption that the essence of entanglement consists in the fact
that the information resides mostly in the correlations then in
properties of individual parts of a composite system \cite{11}.

In fact, the correlations peculiar to entanglement manifest
themselves via local measurements. Since Shannon's definition of
an information channel implies a correlation between input and
output, it is possible to say that an entangled state represents a
virtual information channel between the individual parts of the
system. The main objective of this note is to show that the
maximum entanglement, corresponding to the maximum channel
capacity, is associated with the maximum correlation between the
local measurements, which is equivalent to the maximum uncertainty
of these measurements.

As an example of some considerable interest, we investigate the
case of a composite system, consisting of the two parts with two
quantum degrees of freedom each. Such a system is defined in the
Hilbert space
\begin{eqnarray}
{\cal H}={\bf C}^2 \otimes {\bf C}^2 , \label{1}
\end{eqnarray}
where each subspace is spanned by the basis
\begin{eqnarray}
|e_{1,2} \rangle = | \pm \rangle . \nonumber
\end{eqnarray}
A general state defined in (1) has the form
\begin{eqnarray}
|\psi \rangle = \sum_{i,j=1}^2 a_{ij}|e_i \rangle \otimes |e_j
\rangle \nonumber \\ =a_{11}|++ \rangle +a_{12}|+- \rangle
+a_{21}|-+ \rangle +a_{22}|-- \rangle , \label{2}
\end{eqnarray}
where
\begin{eqnarray}
\langle \psi| \psi \rangle = \sum_{i,j} |a_{ij}|^2=1 . \label{3}
\end{eqnarray}
This state (2) describes the information channel between the two
parts of the system under consideration. It is seen that (2) is
more general than the conventional Schmidt form
\begin{eqnarray}
| \psi_{S} \rangle = \sum_i b_i|e_i \rangle \otimes |e_i \rangle
\label{4}
\end{eqnarray}
usually discussed in the context of entanglement \cite{12}.

In either individual part of the composite system, the local
measurement is represented by the set of infinitesimal generators
of the $SL(2)$ algebra represented by the Pauli matrices
\begin{eqnarray}
\sigma_1 & = & |+ \rangle \langle -|+|- \rangle \langle +|,
\nonumber \\ \sigma_2 & = & -i(|+ \rangle \langle -|-|- \rangle
\langle +|), \label{5} \\ \sigma_3 & = & |+ \rangle \langle +|-|-
\rangle \langle -|. \nonumber
\end{eqnarray}
The physical realization of these operators depends on the
concrete definition of the system under consideration.

It is then clear that the correlation between the local
measurements at different individuals, described by the
correlation function
\begin{eqnarray}
\langle \sigma^{(1)}_{\ell}; \sigma^{(2)}_{\ell'} \rangle \equiv
\langle \sigma^{(1)}_{\ell} \sigma^{(2)}_{\ell'} \rangle - \langle
\sigma^{(1)}_{\ell} \rangle \langle \sigma^{(2)}_{\ell'} \rangle ,
\label{6}
\end{eqnarray}
achieves the maximum at the following condition
\begin{eqnarray}
\forall i, \ell \quad \langle \sigma^{(i)}_{\ell} \rangle =0.
\label{7}
\end{eqnarray}
The averaging in above formulas is performed over the state (2).

In turn, the uncertainty of local measurement described by the
variance
\begin{eqnarray}
\langle \left( \delta \sigma^{(i)}_{\ell} \right)^2 \rangle =
\langle \left( \sigma^{(i)}_{\ell} \right)^2 \rangle - \langle
\sigma^{(i)}_{\ell} \rangle^2 \label{8}
\end{eqnarray}
also achieves the maximum under the same condition (7).

We now note that the opposite condition of minimum uncertainty of
the $SL(2)$ generators corresponds to the so-called atomic
coherent states have been discussed in \cite{13}. It is known that
the coherent states describe almost classical behavior of the
corresponding system. Therefore, all one can expect is that the
states with maximum uncertainty should describe the pure quantum
behavior of the system.

In fact, the condition (7) together with the normalization
condition (3) impose on the matrix
\begin{eqnarray}
A=||a_{ij}|| \label{9}
\end{eqnarray}
of the coefficients in (2) the following constraints
\begin{eqnarray}
|a_{11}|^2+|a_{12}|^2= \frac{1}{2}, \quad |a_{22}|=|a_{11}|, \quad
|a_{21}|=|a_{12}|, \label{10} \\ \arg a_{11}+ \arg a_{22}- \arg
a_{12}- \arg a_{21} = \pm \pi + 2k \pi , \label{11}
\end{eqnarray}
where $k=0,1, \cdots$. These conditions specify an infinite subset
of states (2). In particular, the choice of either $a_{11}=0$ or
$a_{12}=0$ in (10) lead to the two sets of mutually orthogonal EPR
states
\begin{eqnarray}
| \psi_{EPR} \rangle = \frac{1}{\sqrt{2}} (|+- \rangle + e^{i
\phi} |-+ \rangle ) \nonumber
\end{eqnarray}
and
\begin{eqnarray}
| \varphi_{EPR} \rangle = \frac{1}{\sqrt{2}} (|++ \rangle +e^{i
\phi'}|-- \rangle ) , \nonumber
\end{eqnarray}
respectively. Here
\begin{eqnarray}
\phi = \arg a_{21}- \arg a_{1,2} , \nonumber \\ \phi' = \arg
a_{22}- \arg a_{11} . \nonumber
\end{eqnarray}
Another example is provided by the orthogonal states
\begin{eqnarray}
| \psi \rangle & = & \frac{1}{2} [i(|++ \rangle +|-- \rangle )+
|+- \rangle +|-+ \rangle ], \nonumber \\ | \psi' \rangle & = &
\frac{1}{2} [|++ \rangle +|-- \rangle +i(|+- \rangle +|-+ \rangle
)]. \nonumber
\end{eqnarray}

By definition, the states (2) are nonseparable and hence
entangled. Let us now show that the condition (7) implies the
maximum entanglement of the states (2). For this, we consider a
standard way based on the use of reduced density matrices
\begin{eqnarray}
\rho^{(i)} = Tr_i \rho , \quad \quad \rho = | \psi \rangle \langle
\psi | . \label{12}
\end{eqnarray}
It is easily seen that
\begin{eqnarray}
\rho^{(1)} = AA^+, \quad \quad \rho^{(2)} = A^+A. \nonumber
\end{eqnarray}
It is a straightforward matter to arrive at the conclusion that
the conventional conditions of internal consistency of quantum
mechanics \cite{2,3} together with the condition of maximum of
reduced entropy
\begin{eqnarray}
\forall i \quad S_i=-Tr( \rho^{(i)} \ln \rho^{(i)} )= \ln 2
\nonumber
\end{eqnarray}
yield, on allowing of (3), the constraints (10) and (11). In other
words, they are equivalent to the condition (7).

It is no wonder that the requirements of maximum uncertainty of
local measurements and maximum of reduced entropy lead to the same
condition (7). In fact, the group of symmetry of the system under
consideration is
\begin{eqnarray}
G=SU(2) \otimes SU(2) \label{13}
\end{eqnarray}
and $SU(2) \subset SL(2)$. Since $SL(2)$ represents the
complexification of $SU(2)$,  the Pauli operators (5) are
responsible for the dynamical symmetry in the system under
consideration. By means of the transformation
\begin{eqnarray}
g=g_1 \otimes g_2, \quad \quad g \in G \nonumber
\end{eqnarray}
applied to the matrix (9) we get
\begin{eqnarray}
A \rightarrow g_1Ag_2^+, \quad A^+ \rightarrow g_2Ag_1^+.
\label{14}
\end{eqnarray}
The only quantity, which has the physical meaning for the virtual
information channel described by the state (2), is represented by
the invariant with respect to (13), that is, by the trace of
matrices $AA^+$ and $A^+A$:
\begin{eqnarray}
I= Tr(AA^+) \equiv Tr \rho^{(1)}=Tr(A^+A) \equiv Tr \rho^{(2)}.
\nonumber
\end{eqnarray}
Use of (14) allows this to be cast into the form
\begin{eqnarray}
I=Tr(g_1AA^+g_1^+)=Tr(g_2A^+Ag_2^+), \label{15}
\end{eqnarray}
which completes the prove.

In particular, the equality (15) can be used to show that the
Pauli matrices (5) commute with the reduced density matrices
\begin{eqnarray}
\forall i, \ell \quad [ \sigma_{\ell}, \rho^{(i)}]=0, \nonumber
\end{eqnarray}
if the elements of (9) obey the condition (7).

Let us stress that the obtained results for the channel defined by
the relations (2) and (7) are compatible with the quantum
counterpart of Shannon's coding theorem developed by Holevo
\cite{14}.

Similar results can be established for the general GHZ states as
well. In this case, the system is defined in the space
\begin{eqnarray}
{\cal H}_3={\bf C}^2 \otimes {\bf C}^2 \otimes {\bf C}^2
\label{16}
\end{eqnarray}
and the group of symmetry has the form
\begin{eqnarray}
G'=SU(2) \otimes SU(2) \otimes SU(2). \label{17}
\end{eqnarray}
A general state defined in (16) has the following form
\begin{eqnarray}
| \psi \rangle = \sum_{i,j,k=1}^2 x_{ijk}|e_i \rangle \otimes |e_j
\rangle \otimes |e_k \rangle , \quad |e_{1,2} \rangle =| \pm
\rangle , \label{18}
\end{eqnarray}
and obeys the normalization condition. It is straightforward to
show that condition (7) with averaging over the state (18)
provides the maximum of reduced entropy as above. In particular,
condition (7) defines the conventional GHZ states
\begin{eqnarray}
| \psi_{GHZ} \rangle = \frac{1}{\sqrt{2}} (|+++ \rangle \pm |---
\rangle ). \nonumber
\end{eqnarray}
Besides that, there are infinitely many other maximum entangled
states defined by the relations (18) and (7). A nontrivial example
is provided by
\begin{eqnarray}
| \psi \rangle = \frac{1}{\sqrt{8}} [|+++ \rangle +|--- \rangle
-i(|++- \rangle +|--+ \rangle) \nonumber \\ +|+-+ \rangle +|-+-
\rangle +i(|-++ \rangle +|+-- \rangle )]. \nonumber
\end{eqnarray}
In general, the maximum entangled states of an arbitrary
multi-component system can be determined by the condition (7) in
the similar fashion.

To conclude, we have shown that the maximum entanglement
corresponds to the maximum correlation between the local
measurements and to the maximum uncertainty of these measurements.
The quantitative criterion is represented by the condition (7)
which enables us to fairly simplify the analysis of entangled
states. The novelty is that the condition of maximum entanglement
is defined directly in terms of what can be measured for the
individual parts of the system.

Our consideration so far have applied to the composite systems
with individuals defined in the two-dimensional space. As a matter
of fact, the increase of the number of quantum degrees of freedom
per individual part leads to a certain change of local
measurement. For example, the case of free degrees of freedom per
component assumes that the measurement is represented by the eight
Hermitian operators associated with the generators of the $SU(3)$
algebra. This operators are similar, in a certain sense, to the
Stokes operators, describing the polarization of multipole photons
\cite{15}. The maximum  entangled states, in this case, can be
defined in direct analogy to the above result by the requirement
of maximum uncertainty of local measurements. The detailed
investigation of this case deserves special consideration.

\end{document}